\documentclass[10pt,twocolumn,conference]{IEEEtran}

\IEEEoverridecommandlockouts

\usepackage{graphicx}
\usepackage{amsmath}
\usepackage{amssymb}
\usepackage[caption=false]{subfig}
\usepackage[noadjust]{cite}
\usepackage{float}
\usepackage{algorithm}
\usepackage{bibentry}
\usepackage{balance}
\usepackage{algorithm}
\usepackage{algorithmicx}
\usepackage{algpseudocode}
\usepackage{xcolor}
\usepackage{subfig}

\usepackage{url}

\graphicspath{{img/}}

\begin{document}

\bstctlcite{IEEEexample:BSTcontrol}

\title{National Radio Dynamic Zone Concept with Autonomous Aerial and Ground Spectrum Sensors\thanks{This research is supported in part by the NSF award CNS-1939334 and its supplement for studying NRDZs. The authors would like to thank M. Rogers from Carolina Unmanned Vehicles for allowing to reuse Fig.~\ref{fig:Helikite} in this paper.}}

\author{S. J. Maeng, \.{I}. G\"{u}ven\c{c}, M. L. Sichitiu, B. Floyd,\ R. Dutta, T. Zajkowski, O. Ozdemir, and M. Mushi\\
Department of Electrical and Computer Engineering, North Carolina State University, Raleigh, NC\\
{\tt \{smaeng,iguvenc,mlsichit,bafloyd,rdutta,tjzajkow,oozdemi,mjmushi\}@ncsu.edu}}

\maketitle




\begin{abstract}
National radio dynamic zone (NRDZs)  are intended to be geographically bounded areas within which controlled experiments can be carried out  while protecting the nearby licensed users of the spectrum. An NRDZ will facilitate research and development of new spectrum technologies, waveforms, and protocols, in typical outdoor operational environments of such technologies. 
In this paper, we introduce and describe an NRDZ concept that relies on a combination of autonomous aerial and ground sensor nodes for spectrum sensing and radio environment monitoring (REM). 
We elaborate on key characteristics and features of an NRDZ to enable advanced wireless experimentation while also coexisting with licensed users. 
Some preliminary results based on simulation and experimental evaluations are also provided on out-of-zone leakage monitoring and real-time REMs.
\end{abstract}



\section{Introduction} \label{sec:intro}
 
National radio dynamic zones (NRDZs)  are recently being conceptualized in the United States as geographical areas where transmitters and receivers can operate while protecting the nearby licensed users of the spectrum. Designing, developing, and testing new advanced wireless technologies requires access to typical operational environments of such technologies for extensive testing. While NRDZs can serve as key enablers to facilitate such experimentation, how to design, deploy, and operate an NRDZ is a complex problem. Most critically, real-time spectrum monitoring across the NRDZ geographical area is of paramount important to ensure that nearby passive and existing users are protected from interference all the time. 

While the United States does not have any NRDZ at this time, there are currently two National Radio Quiet Zones (NRQZs): one is in Green Bank, West Virginia~\cite{WV_NRQZ}, the other is in Table Mountain, Colorado~\cite{CO_NRQZ}. An NRQZ is a geographical area where there are special rules (established by the National Telecommunications and Information Administration (NTIA)) to protect special receivers inside the NRQZ area, such as sensitive radars, from other radio signal sources, such as cellular or TV broadcast signals, located outside of that area. On the other hand, an NRDZ would protect normal receivers outside the zone from special transmitters inside the zone~\cite{NRDZ_vs_NRQZ}. The special transmitters range from the directed energy systems to high-power microwave transmitters. The prime criteria to assess an NRDZ would be how much power may escape outside of the NRDZ boundary. In addition to protecting some active users of the spectrum, an NRDZ should protect sensitive passive users nearby whose interference thresholds are orders of magnitude lower than those of active users. 

A critical question regarding the feasibility of NRDZs is how effectively one can predict, detect, and prevent leakage of radio emission from an NRDZ to specific receivers or the specific geographical region. In particular, the highly sensitive radio-astronomy passive receivers and the NRDZ's step-sisters, the NRQZs, are highlighted as the most concerned ones. Besides, any number of collocated incumbent receivers may have to be efficiently protected. 

In this paper, we present our views on critical design  challenges for NRDZs, and  we present  an NRDZ concept with aerial and ground mobile sensors for effective spectrum monitoring. We subsequently discuss research and experimentation challenges for such an NRDZ framework. 
We discuss that Aerial Experimentation and Research Platform for Advanced Wireless) AERPAW~\cite{marojevic2020advanced}, a recent testbed platform that offers various types of experimentation with vehicular and aerial fixed and mobile nodes, can help in developing, testing, and evaluating technical concepts for future NRDZs with autonomous ground and aerial spectrum sensors. We  provide preliminary theoretical and experimental results on real-time REM development and spectrum monitoring, and elaborate on future NRDZ experiment concepts that can be carried over AERPAW.

\section{Features and Requirements of an NRDZ} \label{sec:}

In this section, we provide our views on various essential features, requirements, and characteristics of an NRDZ.

\subsection{Geographical Area for an NRDZ}

NRDZs can be used to test experimental radio technologies such as 5G/6G wireless systems, high-power directional transmissions, and new antenna concepts with complex radiation patterns, and to evaluate propagation characteristics in frequency bands outside of the use cases that those bands have been traditionally used. As such, it is critical to limit the leakage of electromagnetic radiation to sensitive receivers outside of the NRDZ. The geographical area, terrain characteristics, and types of sensitive receivers nearby an NRDZ are hence play key roles on the design of the NRDZ. 

In some cases, it can be possible to know the locations of receivers to be protected, and NRDZ can aim to limit the power towards those known receivers outside of the zone. Although NRDZs may be constructed in electromagnetically impenetrable zones conceptually (e.g., underground, deep in the ocean, or inside a Faraday cage), typical outdoor NRDZs may potentially impact sensitive receivers that may be thousands of square miles away. The specific location, size, and boundary of an NRDZ should therefore be very carefully designed, taking into consideration the specific locations of receivers that may be potentially impacted from the experiments in the NRDZ, their specific frequencies, receiver sensitivities, and propagation characteristics. For example, a geographical area that may be suitable for NRDZ experiments in one specific frequency may not be appropriate in some other frequency due to nearby sensitive receivers in the latter. 

\subsection{Supported Frequency Bands in an NRDZ}
We have a thorough understanding of the fundamental propagation properties of electromagnetic radiation for low-band and mid-band frequencies (below $6$~GHz)  that have been used for a long period. Therefore, after careful design and evaluation of impacts to nearby receivers, these frequencies may be more conveniently used for NRDZ experimentation. It is recently shown that higher frequencies, such as millimeter wave (mmWave) and terahertz (THz), propagate better than it was previously believed~\cite{rappaport2013millimeter}. These frequencies offer cellular coverage in urban environments as well where reflection and diffraction effects are more likely. 

While beamforming techniques that utilize a larger number of antenna elements in the same physical antenna area at higher radio frequencies can help extend the range further, it is particularly troublesome for an NRDZ, which seeks the exact opposite of coverage to protect sensitive receivers. We take into account of mmWave and THz models for the geography of the NRDZ as well, in order to ensure the existing receivers in those bands are being protected from leakage outside of the  NRDZ. For directional transmissions, the incidence angle to a sensitive receiver also matters, and it may allow some flexibility for NRDZ experiments and nearby receivers to coexist in the same band. If any coordination may be allowed and possible among an NRDZ and the nearby networks, a directional (e.g. mmWave) base-station may choose to schedule its users in a certain combination of  directions and spectrum, in order to avoid interference from transmissions happening in a nearby NRDZ, and vice versa.     

Despite the recent work on air-to-ground signal propagation in mmWave and other bands~\cite{khawaja2019survey}, there is still a need to understand radio propagation to/from aerial transceivers. An NRDZ may need to protect sensitive receivers of non-terrestrial networks (NTNs), hence propagation in bands of interest for such use cases is important to study. While NTN receivers may be existing passive/active users of the band, they can also be aerial sensors deployed by the NRDZ, to evaluate signal leakage outside of the NRDZ. 

\subsection{Radio Environment Maps for Signal Leakage Monitoring}

Accurate propagation models for a given geographical area are critical to find the adequate locations and boundaries of NRDZs and estimate the amount of the leakage of electromagnetic radiation from the NRDZ, in the bands supported by the NRDZ. 
Such models can help in prior planning of transmitter locations, transmission powers, and boundaries of an NRDZ. 
However, no matter how detailed a  model is, it is practically impossible to foresee all possible circumstances that may affect radio propagation in a real experiment. Therefore, an NRDZ experiment should be supported by real-time spectrum monitoring of the ongoing experiment, and necessary actions should be taken to avoid interference to nearby passive/active users of the spectrum. 

Ideally, the RF monitoring system in an NRDZ needs to produce a real-time radio environment map (REM) at all relevant frequencies and locations. To the best of our knowledge, currently, there is no instrument that could produce such a comprehensive and exact map. However, a reasonably accurate approximation of the ideal result can be obtained. For instance, REMs~\cite{yilmaz2013radio} have been used for generating the coverage of cellular networks. Commonly, Kriging interpolation~\cite{sato2017kriging} have been used to transform measurements that are collected sparsely over a geographical region into a continuous-coverage REM, and cooperative~\cite{xue2014cooperative} and crowd-sourced spectrum sensing~\cite{jin2018privacy} have been studied for over a decade. Recently, machine learning-based cooperative spectrum sensing has been proposed in~\cite{thilina2013machine}. There have been several experimental studies as well. In~\cite{atanasovski2011constructing}, a distributed spectrum sensing approach is adopted to generate a real-time REM by utilizing various forms of sensors including USRP. REMs are experimentally studied in resource allocation improvements in WiFi~\cite{dionisio2017experimentation}.




\begin{figure}[t]
	\centering
	\vspace{-0.0in}
    \includegraphics[width=0.35\textwidth]{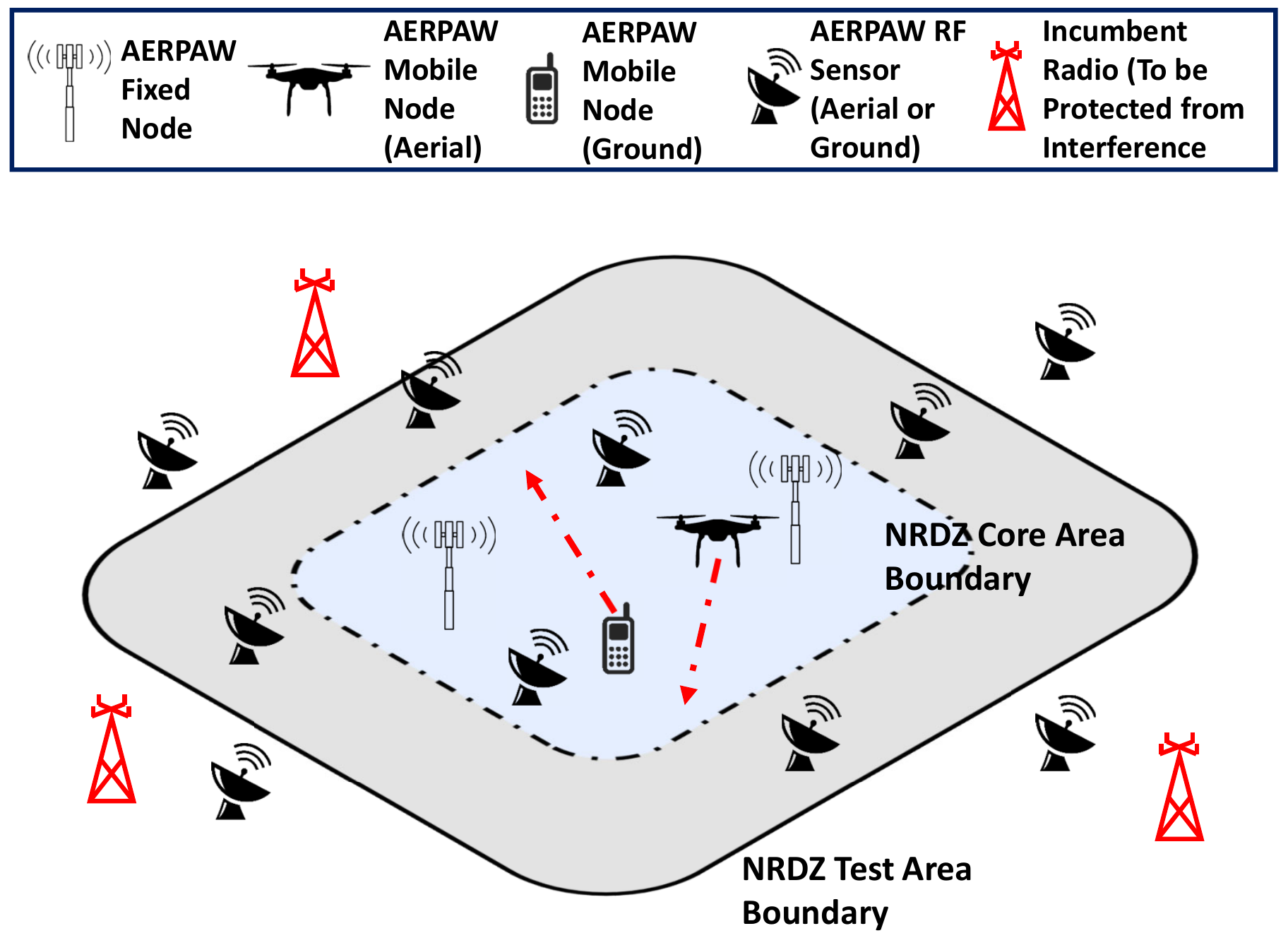}
	\caption{Considered NRDZ concept with test and core areas.}\label{fig:NRDZ_Concept}\vspace{-0.1in}
\end{figure}

\begin{figure}[t]
	\centering
    \includegraphics[width=0.35\textwidth]{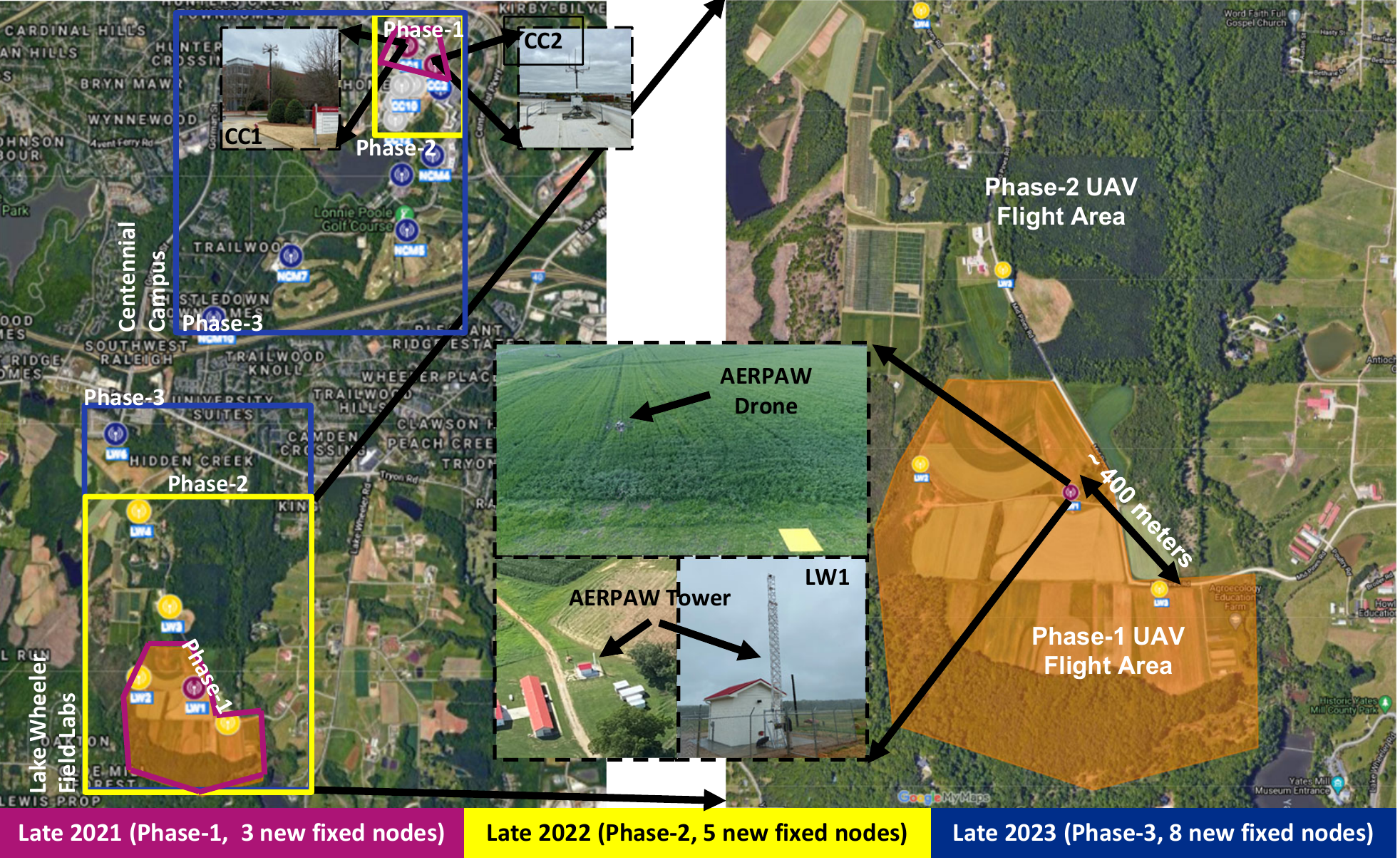}
	\caption{AERPAW fixed node locations and timeline.}\label{fig:AERPAW_map}\vspace{-0.2in}
\end{figure}

\subsection{Hardware and Software Infrastructure}

From the above discussions, it is clear that an NRDZ would consist of several elements: experimental transmitters, receivers, and a set of RF monitors. The software/hardware that could support real-time computing of extensive amount of data collected by the RF monitors is necessary in order to generate an REM. For orchestration of various hardware resources, a management backplane network connecting the various hardware resources is required, ideally with a low-latency high-throughput optical fiber network for the terrestrial hardware, and a stable cellular connection for the mobile/aerial parts of the system that utilizes spectrum controlled (at least locally) by the NRDZ facility itself.  

Additionally, for the ability to use arbitrary subsets of such resources on an agile, flexible basis, it is desirable that network be configurable to connect and to isolate arbitrary subsets of the hardware connected to the network, simultaneously.  Ideally, a software defined network (SDN) based programmable approach, utilizing cross-layer orchestration of access control, would be used for control of the backplane network, and be able to seamlessly integrate the cellular connections into the mobile/aerial components. 

\section{Conceptualizing  NRDZs}



In Fig.~\ref{fig:NRDZ_Concept}, we illustrate our  high-level concept of an NRDZ that consists of fixed and mobile spectrum monitoring sensors, transmitters, and receivers. As an example infrastructure for developing and testing early NRDZ concepts, we consider the National Science Foundation (NSF) AERPAW platform located at NC State University~\cite{FCC_innovation_zone,marojevic2020advanced,chowdhury2021taxonomy} (see Fig.~\ref{fig:AERPAW_map}), that consists of programmable radios, RF monitors, control software capable of leveraging the hardware, as well as aerial and ground vehicles that can serve as mobility anchors for portable wireless equipment and sensors.  

The signal sources (e.g., fixed and mobile AERPAW nodes) are located only within the NRDZ core area (NRDZ-CA) and all the incumbent (passive/active) receivers (IPARs) that need to be protected, illustrated in red, are located outside of the NRDZ test area (NRDZ-TA). In particular, these receivers need to be protected from any leakage power from the NRDZ. The region between the boundary of the NRDZ-TA and the NRDZ-CA is allocated as an NRDZ guard area (NRDZ-GA) where no transmission sources and receivers are allowed. Various RF monitoring sensors (RFMS) need to be deployed by an NRDZ inside and outside of its NRDZ-TA. The RFMSs can be either fixed or mobile, and can be on the ground or at an aerial vehicle (e.g., drone, helikite).

Considering this NRDZ concept, there are several design and research challenges. For example, given a specific frequency band, the propagation characteristics for the operational environment, transmit powers, and receiver sensitivities, the dimensioning of the NRDZ-TA, NRDZ-CA, and NRDZ-GA each need to be studied. Heights and directionalities of the NRDZ transmitters/receivers and IPARs (e.g., with aerial transmitters/receivers) will also affect the NRDZ coverage. The NRDZ-GA and NRDZ-CA should be large enough  to allow meaningful experiments, while also introducing a sufficient physical separation between the NRDZ transmitters and the IPARs.  Note that in some cases, IPARs may be at high altitude, such as satellite receivers, in which case the NRDZ-TA, NRDZ-CA, and NRDZ-GA concepts may need to be expanded to the 3D space, and aerial spectrum sensors then need to characterize spectrum leakage in the 3D space.

As an alternative to the concept of NRDZ-GA in Fig.~\ref{fig:NRDZ_Concept}, one may consider exclusion zones around sensitive IPARs. This may depend on the number,  locations, and sensitivities of IPARs around the NRDZ. For example, a highly sensitive passive receiver may require its own exclusion zone around it where no NRDZ transmitters are allowed. Cellular networks, as an example, may have hundreds of fixed and mobile nodes around the vicinity of an NRDZ, and it may be impractical to consider the use of exclusion zones. 
This may require NRDZ to have mechanisms for actively detecting and monitoring in-coverage IPARs and eliminating interference to them, e.g. by changing the transmission bands, beam directions, and/or transmit power levels. 

\section{Future Research Directions for NRDZs}

In this section, we identify several research directions that may help in enabling NRDZs in the future. 

\subsection{Spectrum Monitoring and Signal Leakage Sensing}

There are several interesting questions to study  how the RFMSs can be deployed and operated in an NRDZ. For example, detection of the signal leakage outside of the NRDZ-CA is a major challenge, and should be identified and localized by the RFMS network. In this context, the specific density and locations of the RFMS sensors in the NRDZ-CA, NRDZ-GA, and outside of the NRDZ-TA, needs to be researched, for enabling a real-time REM system. How can techniques such as beamforming and intelligent reflectors be properly used to prevent leakage from the NRDZ-TA should be further studied. Especially when an NRDZ involves mobile transmitters, there will be additional challenges for dynamic monitoring of the spectrum use and signal leakage, which requires further studies. In this context, aerial and ground mobile RFMS sensors can be designed with artificial intelligence (AI) based trajectory updates for keeping the REM up to date and accurate. 

In addition to spectrum monitoring for sub-6 GHz, mmWave software-defined arrays (SDAs) can be used to sense emissions from mmWave signals sources.  The receiver can measure the spectrum from both in-band and out-of-band (OOB) emissions. For example, 28 GHz phased arrays still work at 24 GHz (where passive sensitive receivers may be present), with somewhat compromised beam steering and RF signal processing. Above all, the proper characterization, calibration, and control of the array in OOB are most significant, since the phase and amplitude control in OOB could be quite different in order to steer beams accurately. We can utilize code-modulated interferometry (CMI) for this purpose~\cite{GRE18,HON19,CHA19}. In CMI, all elements in the array are modulated using unique codes and then the vector responses of all phase shifters in free space can be extracted in parallel using a simple scalar detector together with digital signal processing, which allows us to calibrate the array at any frequency.


\begin{figure}[h]
	\centering
	\subfloat[Keysight RF sensor.]{ \includegraphics[width=0.17\textwidth]{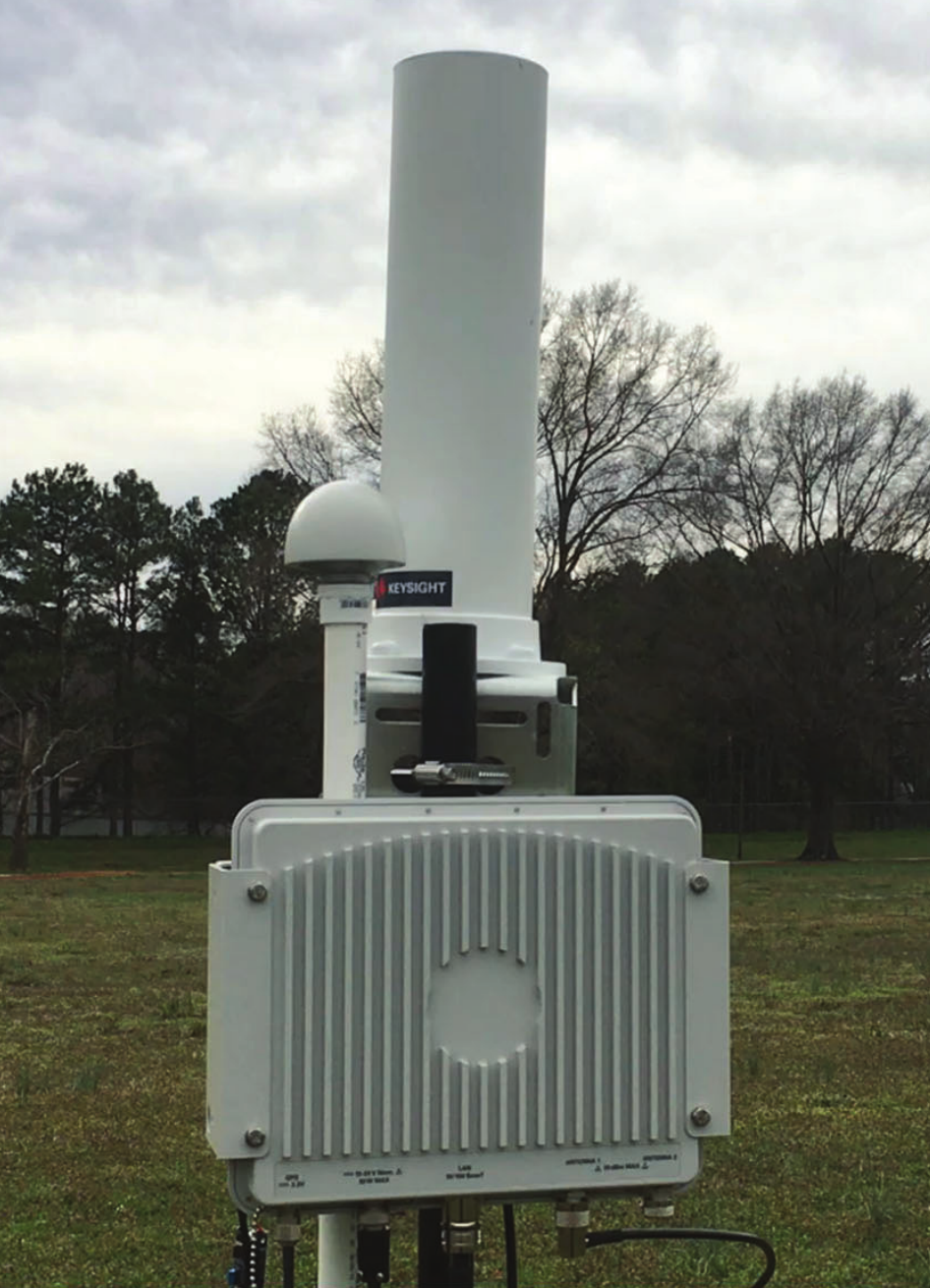}}~~
	\subfloat[Tracking drones.]{ \includegraphics[width=0.27\textwidth]{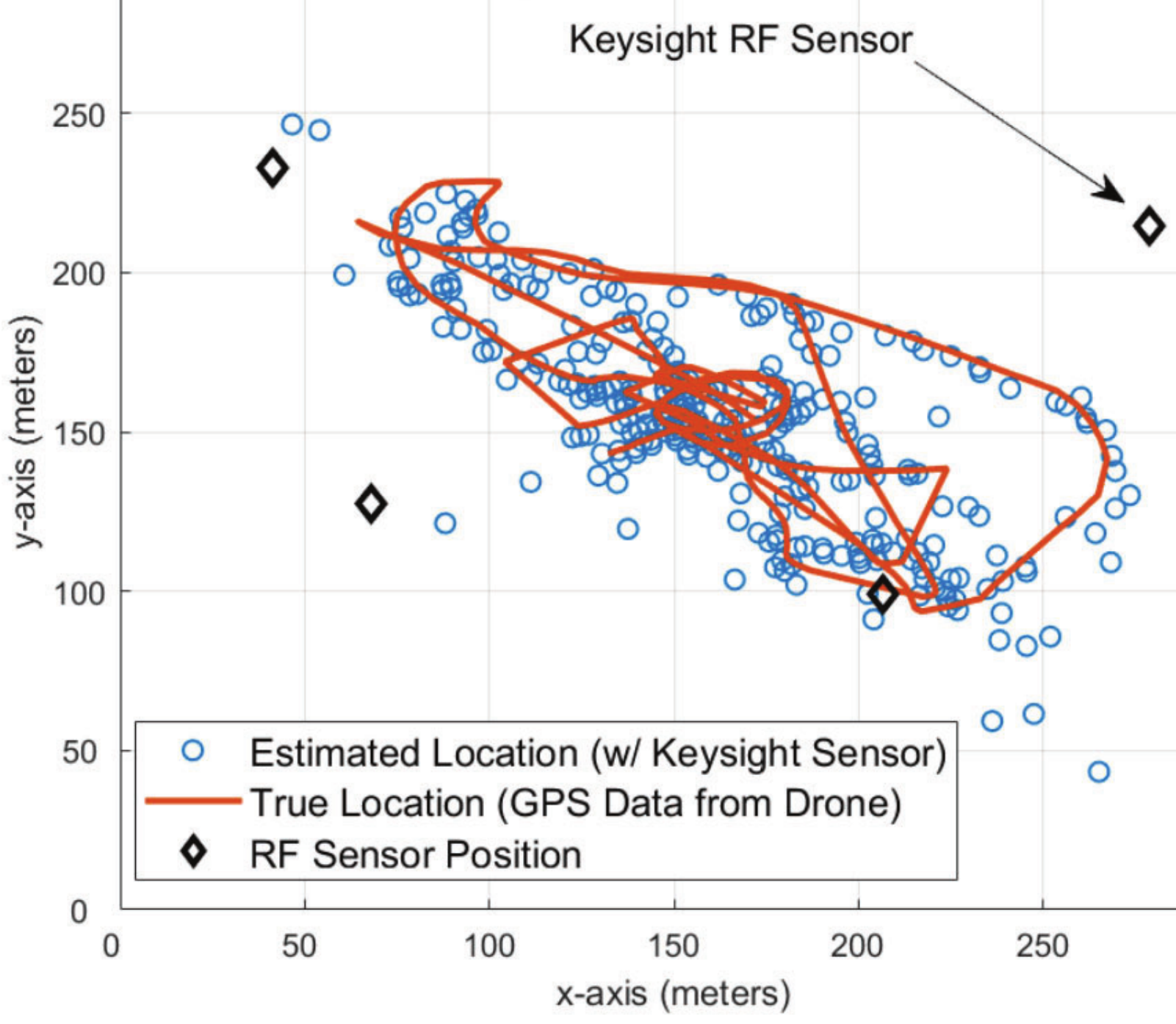}}
	\caption{Localizing a mobile RF signal source using Keysight RF sensors.}\label{fig:drone_localization}\vspace{-0.2in}
\end{figure}

\subsection{Signal and Interferer Source Tracking}

In addition to supporting a real-time REM, we can also localize sources inside the NRDZ-CA that have the potential to interfere with the IPARs. In order to localize and track signal sources, accurate recognition and classification of such signal sources by the RFMS network is critical. To achieve this, we have done preliminary work with Keysight RF sensors, which inherently provide time-difference of arrival (TDOA) based source localization due to precise synchronization between the sensors. They also have built-in signal classification features; for example, one can train the sensors with existing LTE, WiFi, UAV remote control, among other signals, and subsequently, the sensors will be able to recognize the signal class observed at a given frequency in real-time operations. 

Fig.~\ref{fig:drone_localization} shows preliminary results of localization experiments in Dorothea Dix Park in Raleigh, NC, from experiments carried out in March 2020~\cite{bhattacherjee2021experimental}. We deployed four Keysight RF sensors on tripods, and accurately detected and classified the 2.4~GHz signals from multiple drones. The RFMS network was then able to localize the drone signal source by TDOA with a median of error smaller than 10~m. Additional filtering techniques, such as Kalman filters and particle filters, can further improve localization/tracking performance as show  in~\cite{bhattacherjee2021experimental}. This then allows to continuously track the location of potential interference sources and even the locations of IPARs within the NRDZ geographical area. 
Alternatively, the signal sources may be forced to share their GPS location  with the NRDZ system. However, that may not always be feasible, since some mobile nodes may not have GPS, and there may always be third party non-cooperative signal sources. 

\subsection{Propagation Measurements/Modeling with Aerial Nodes}

As discussed in the previous subsection, propagation characteristics in a specific environment and frequency band carry critical importance for designing, deploying, and operating an NRDZ. Such propagation characteristics include but are not limited to time selectivity, frequency selectivity, statistics of angle of arrival/departure of multipath components, signal correlation across multiple antennas, path loss, shadowing, spatial consistency, among others. In this context, we use drones and rovers that travel over pre-defined waypoints for collecting propagation data from ground and aerial transmitters, and to characterize the 3D propagation in the immediate as well as distant neighborhoods of fixed RFMS sensors. 

The radio environment can also be scanned using a LIDAR scanner, and the scanned environment can be transferred to a ray-tracing simulator, for developing realistic propagation models of the same environment. Such models can then be used for pre-planning of NRDZ geographical boundaries and for setting specific constraints on an NRDZ experiment. Such models can also be updated continuously based on new measurements from the field, e.g. to capture variations due to foliage, new construction, weather effects, among others. At mmWave and THz frequencies, the use of absorbers, passive/active meta-surfaces~\cite{ozturk2021channel}, and metallic reflectors~\cite{khawaja2020coverage} can help contain/steer the signals of interest within the NRDZ environment. In addition, mobile aerial and ground platforms that may carry such reflectors can be used to actively contain the signals inside the NRDZ by a dynamic aerial trajectories. More specifically, signal transmissions that may cause harmful interference to outside the NRDZ may be evaded by forming opportunistic scatterers in the 3D space through UAV/UGV trajectory design.

\begin{figure}[t]
	\centering
	\vspace{-0.0in}
	   \subfloat[Helikite from Carolina Unmanned Vehicles~\cite{CUV_Website}.]{
    \includegraphics[width=0.18\textwidth]{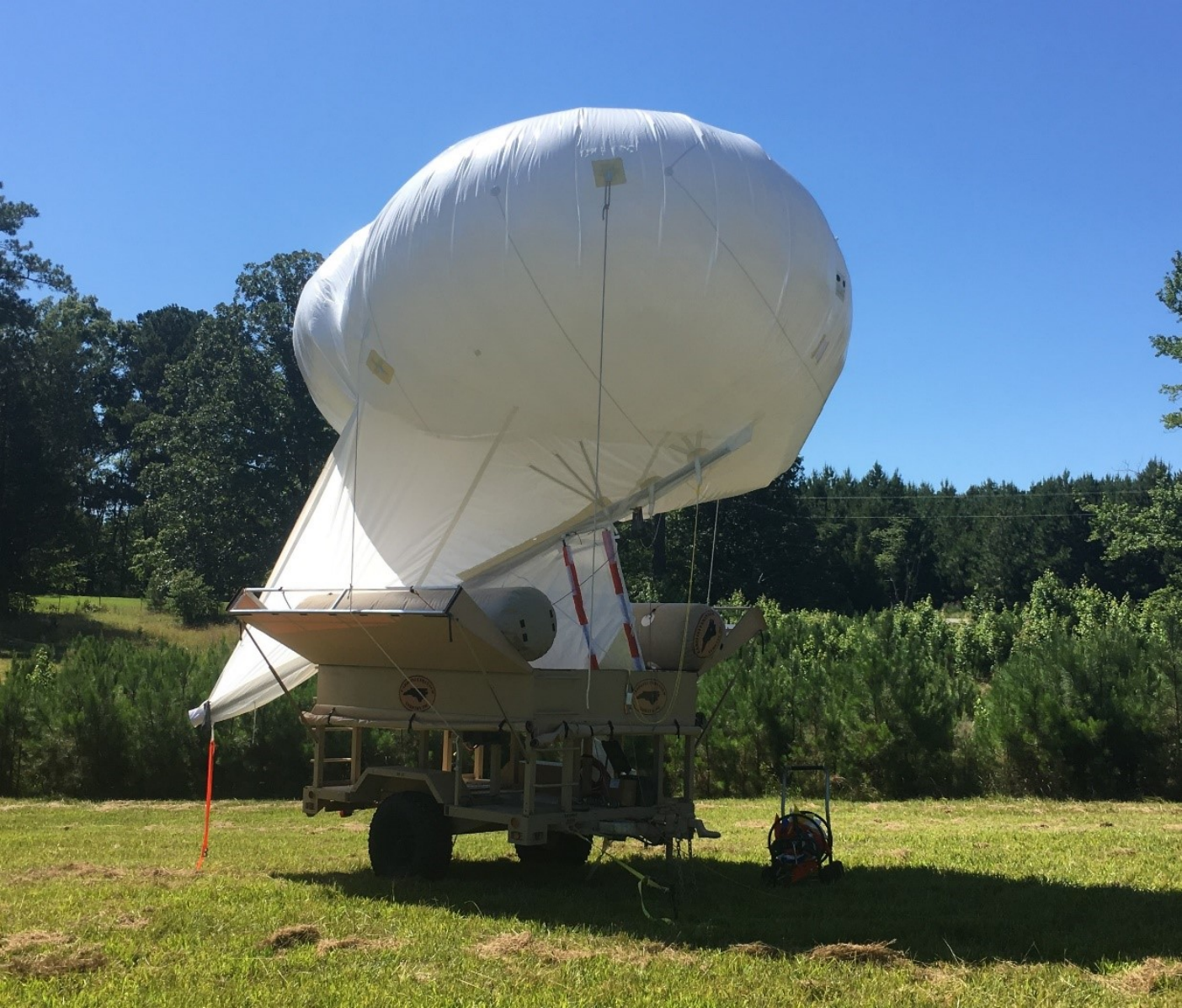}}~~
  \subfloat[LoS range and available payload vs. helikite' height.]{
    \includegraphics[width=0.25\textwidth]{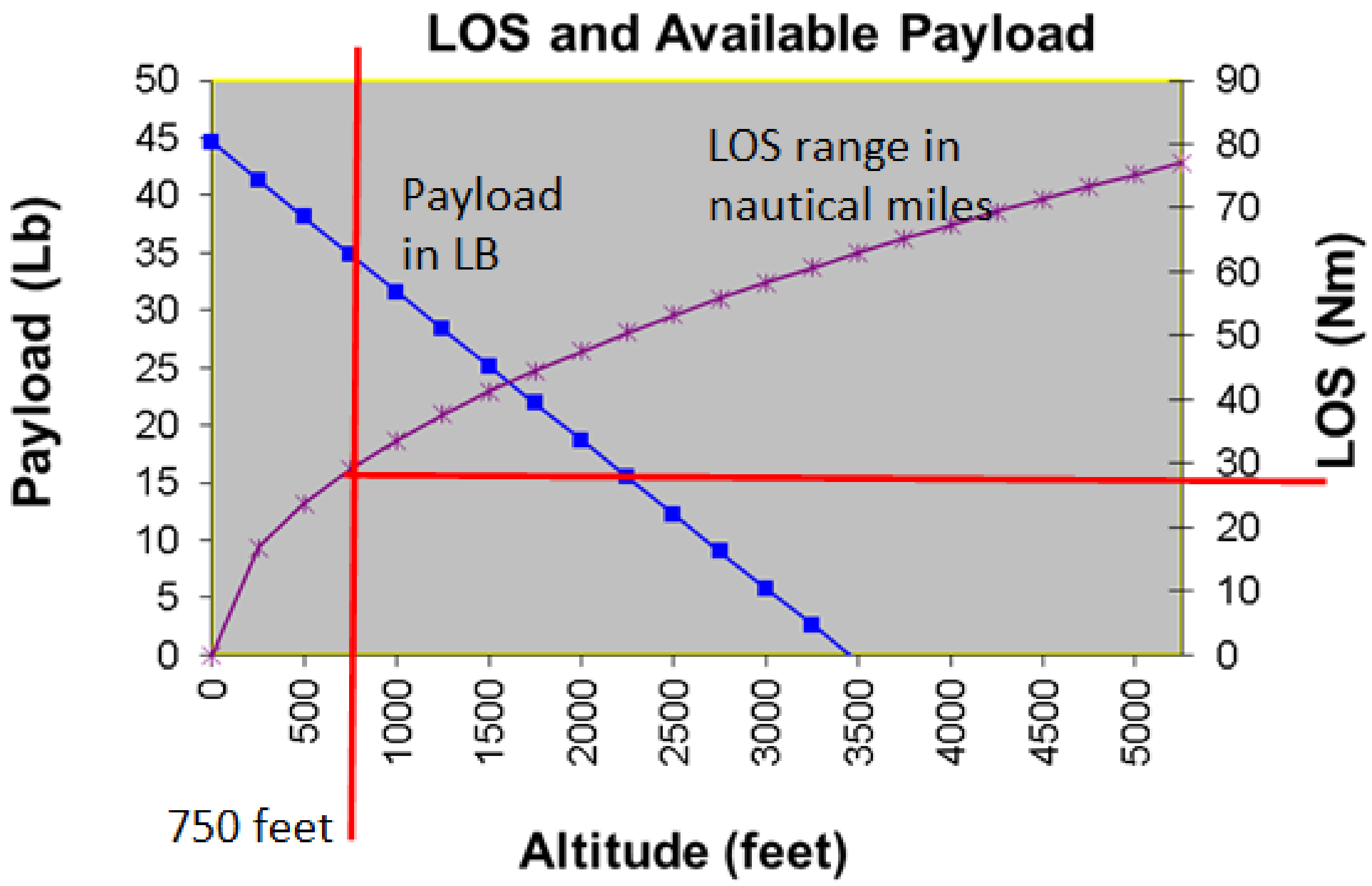}}
	\caption{Helikite deployment at AERPAW.}\label{fig:Helikite}\vspace{-0.3in}
\end{figure}

To support experiments at higher altitudes and longer duration, helikite platforms can be utilized. For example, the AERPAW platform is deploying a helikite platform from Carolina Unmanned Vehicles (see Fig.~\ref{fig:Helikite}a).  In Fig.~\ref{fig:Helikite}b, we present the line-of-sight (LoS) range and available payload in terms of the helikite altitude. The calculation of the LoS range takes into account of the  spherical earth model but does not include the effect of refraction property of the radio propagation, which may affect sub-6 GHz radio propagation. The helikite can carry at least 10 lbs of payload at the maximum operating height of approximately 750 feet. While UAVs can be used for aerial wireless experiments, the current AERPAW UAVs can not fly longer than 40 minutes without a tether. A powered tether also allows long-duration operations for UAVs up to 300 feet. 

In Section~\ref{Sec:PowerLeakageModeling}, we present our early concepts and results for out-of-zone signal leakage monitoring at an NRDZ considering sub-6 GHz. Section~\ref{Section:SpectrumCompliance} will then present preliminary spectrum  measurements from the AERPAW spectrum compliance sensors for frequencies up to 3~GHz.

\begin{figure*}[t]
	\centering
    \subfloat[Proposed system model for NRDZ.]{\includegraphics[width=0.28\textwidth]{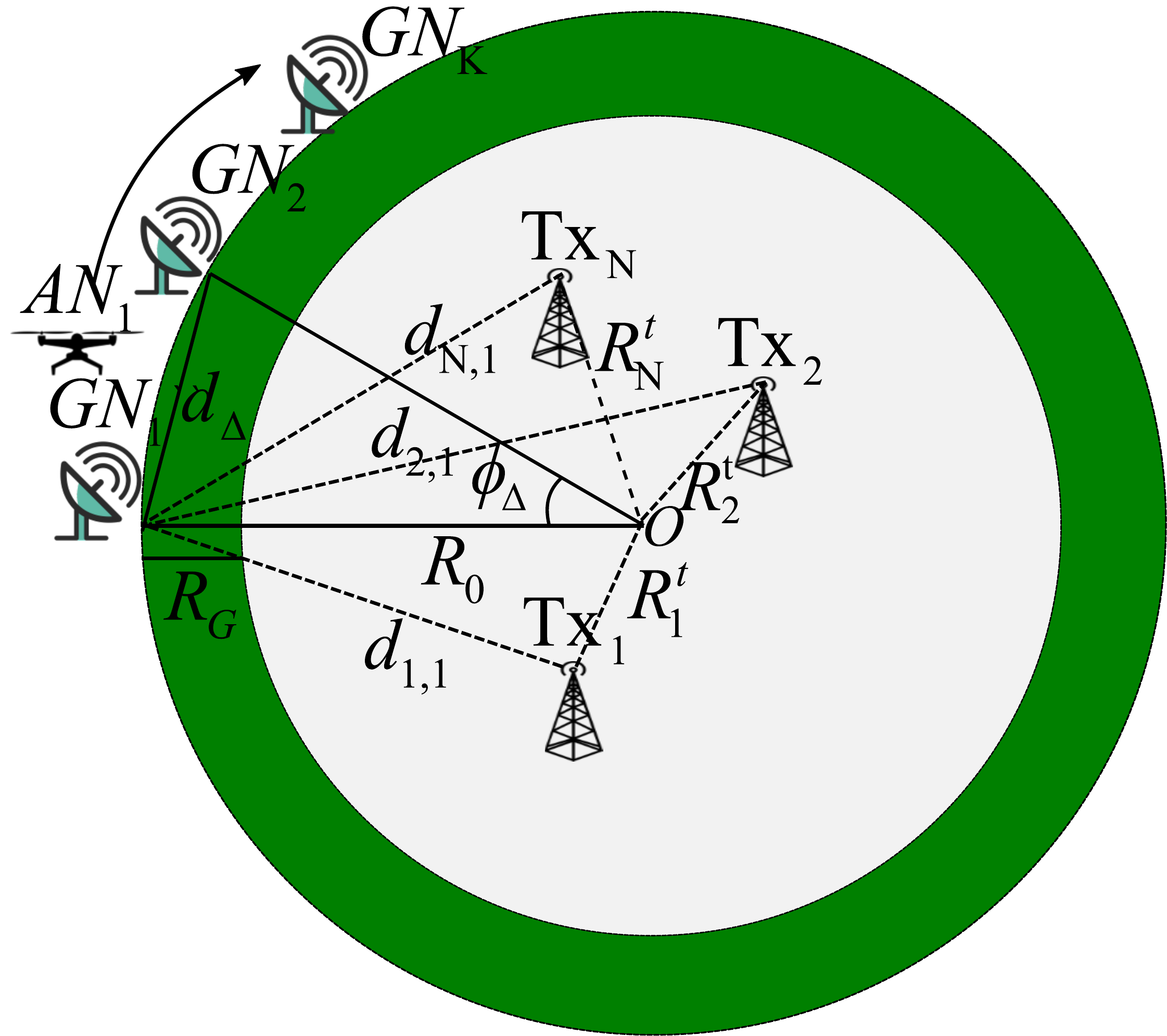}\label{fig:system_model}}~~
    \subfloat[RMSE versus angle space for different $R_0$ $\eta$.]{\includegraphics[width=0.35\textwidth]{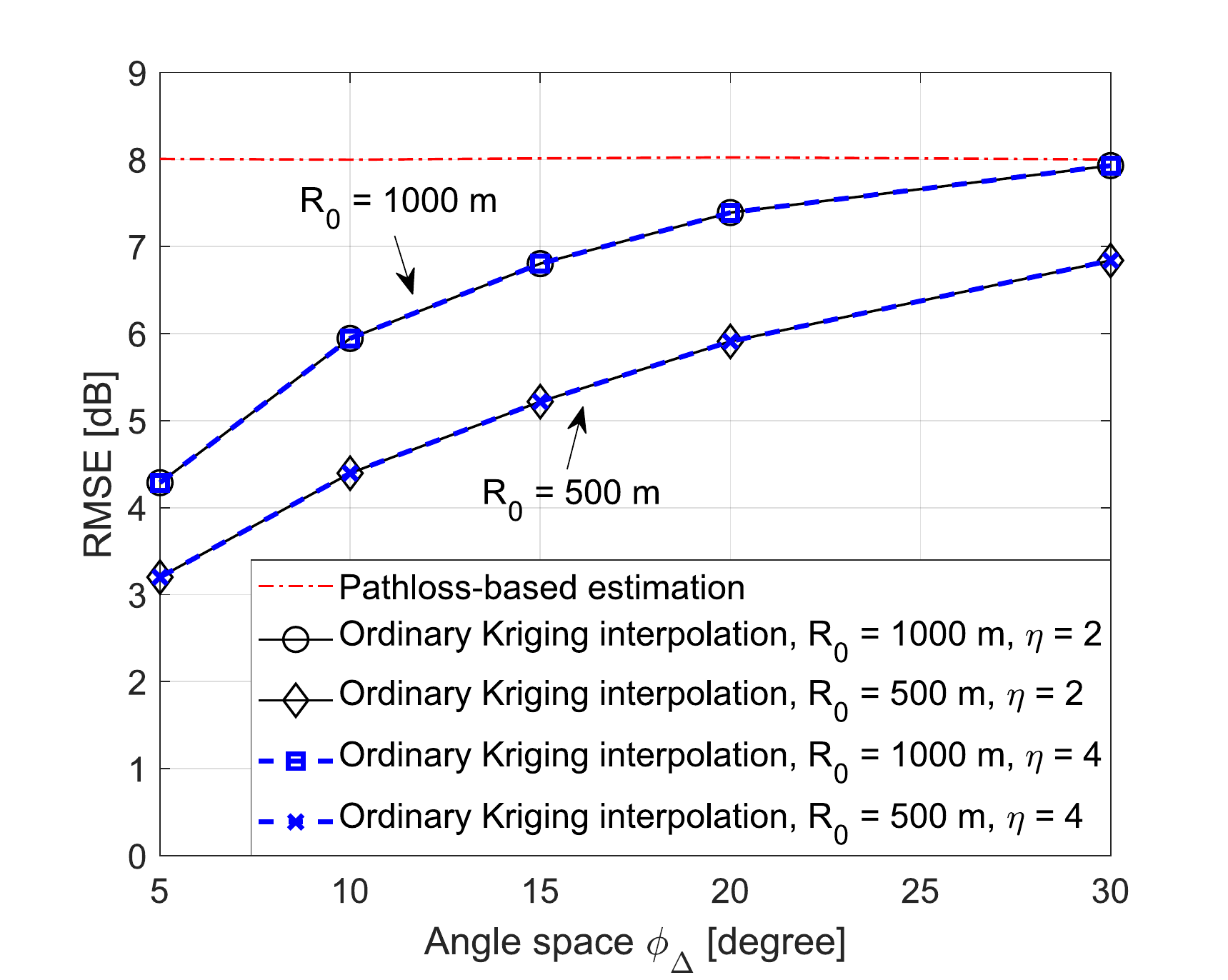}\label{fig:Kriging}}~~
    \subfloat[Spectrum compliance monitoring.]{\includegraphics[width=0.3\textwidth]{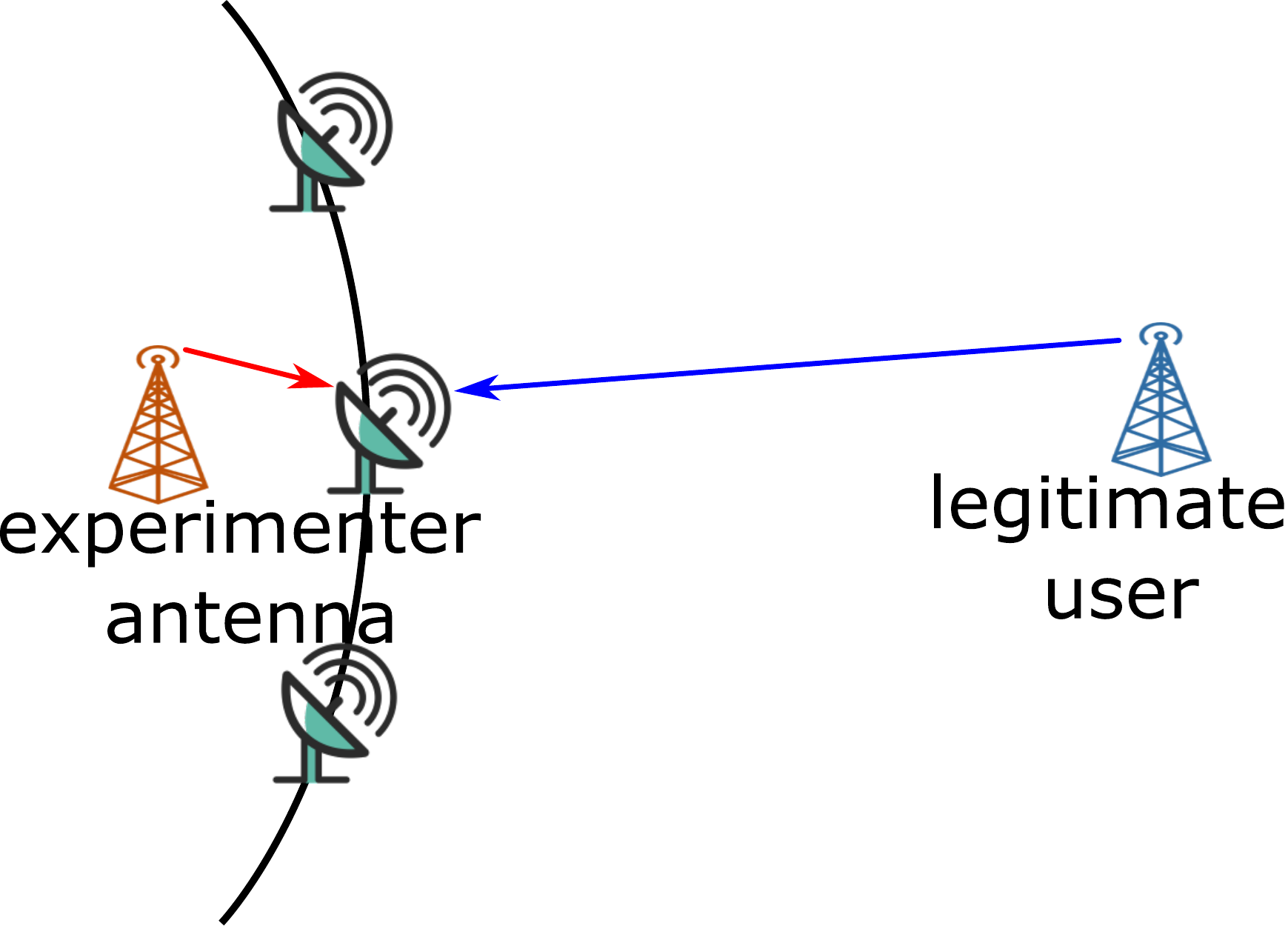}\label{fig:spectrum_compliance}}
	\caption{Power leakage sensing and the spectrum compliance monitoring in NRDZs.}\vspace{-0.25in}
\end{figure*}



\section{Power Leakage Modeling and Sensing}\label{Sec:PowerLeakageModeling}

In this section, we propose a leakage power measurement algorithm to monitor and estimate the signal powers from multiple transmit sources inside of an NRDZ-CA. 
Extending from Fig.~\ref{fig:NRDZ_Concept}, the proposed system model of NRDZ is illustrated in Fig.~\ref{fig:system_model}, which considers a simple disk model for analytical tractability. Fixed nodes and mobile aerial nodes are deployed at the boundary of the NRDZ-TA, at a distance $R_0$, to sense the received signal power from the multiple transmitters. The transmitters are away from the origin of the NRDZ-TA at a distance $R^{\rm t}$, but they are prevented from being located at green zone with a width $R_G$ to protect receivers from the leakage from transmitters. For the sake of the efficiency and the limitation of the number of available nodes, ground and aerial sensing nodes are sparsely positioned with uniform angular space ($\phi_{\Delta}$) and distance ($d_{\Delta}$) between adjacent nodes. The goal of the sensor nodes is to estimate and generate the geographical 3D REM over a dense grid by cooperative interpolation technique. 

\subsection{Channel Model for Multi-Sensor Leakage Sensing}

The received signal strength of sensor nodes is generally modeled by path-loss and shadow fading. Path-loss is decided by the distance between source and receiver, and the signal strength declines  exponentially by the travel distance. However, even if the distance between source and destination is identical, the signal power can be varied by the shadowing effect, which represents the variability of the power around its expected value. The small-scale fading, such as the multi-path effect, can be neglected in the spatial dimensions since it can be averaged out at the time dimension. The shadow fading is commonly modeled by a jointly lognormal distribution. 

If an aerial sensor node moves following a trajectory or two different location fixed sensor nodes receives the signal from the same source, auto-correlation of shadowing between two signals mostly depends on the distance between two locations. On the other hand, the cross-correlation between received signals from the different two sources at the same sensor dominantly depends on the azimuth angle between two sources from the sensor. Besides, it is found that the cross-correlation between two different locations with different sources can be modeled by the combination of two former distance and angle-dependent models with a reasonable assumption of slowly varying cross-correlation.


\subsection{Received Power Interpolation by Kriging}

Thanks to the aforementioned correlation model between signals from different sources and locations, the prediction of the signal at an unknown location is available from the measured signal from the nearby sensors. At first, we need to estimate unknown parameters of both auto-correlation and cross-correlation models. The method of moments and the maximum likelihood method can be used to estimate them. Subsequently, the ordinary Kriging method, which minimizes the mean-squared prediction error, is applied to interpolate the signal power~\cite{sato2017kriging}. 
Our preliminary results in Fig.~\ref{fig:Kriging} show that the proposed interpolation technique by Kriging outperforms path-loss based power estimation. We can also observe that as angular space between the sensors increases, the estimation performance degrades. This means that finding the proper density of sensors is necessary for accurate estimation of the REM. In addition, we can observe that the root mean square error (RMSE) decreases as the NRDZ radius ($R_0$) becomes smaller and the proposed Kriging interpolation is robust to path-loss coefficient ($\eta$). 
Although the proposed algorithm and the results in Fig.~\ref{fig:Kriging} consider only ground sensors, we will extend this interpolation technique to aerial nodes and design a 3D power map in order to protect ground receivers as well as aerial passive receivers. 


    

    
    

\begin{figure}[!t]
	\centering
	\vspace{-0.0in}
    \subfloat[Frequency band 100 - 500 MHz]{\includegraphics[width=0.5\columnwidth]{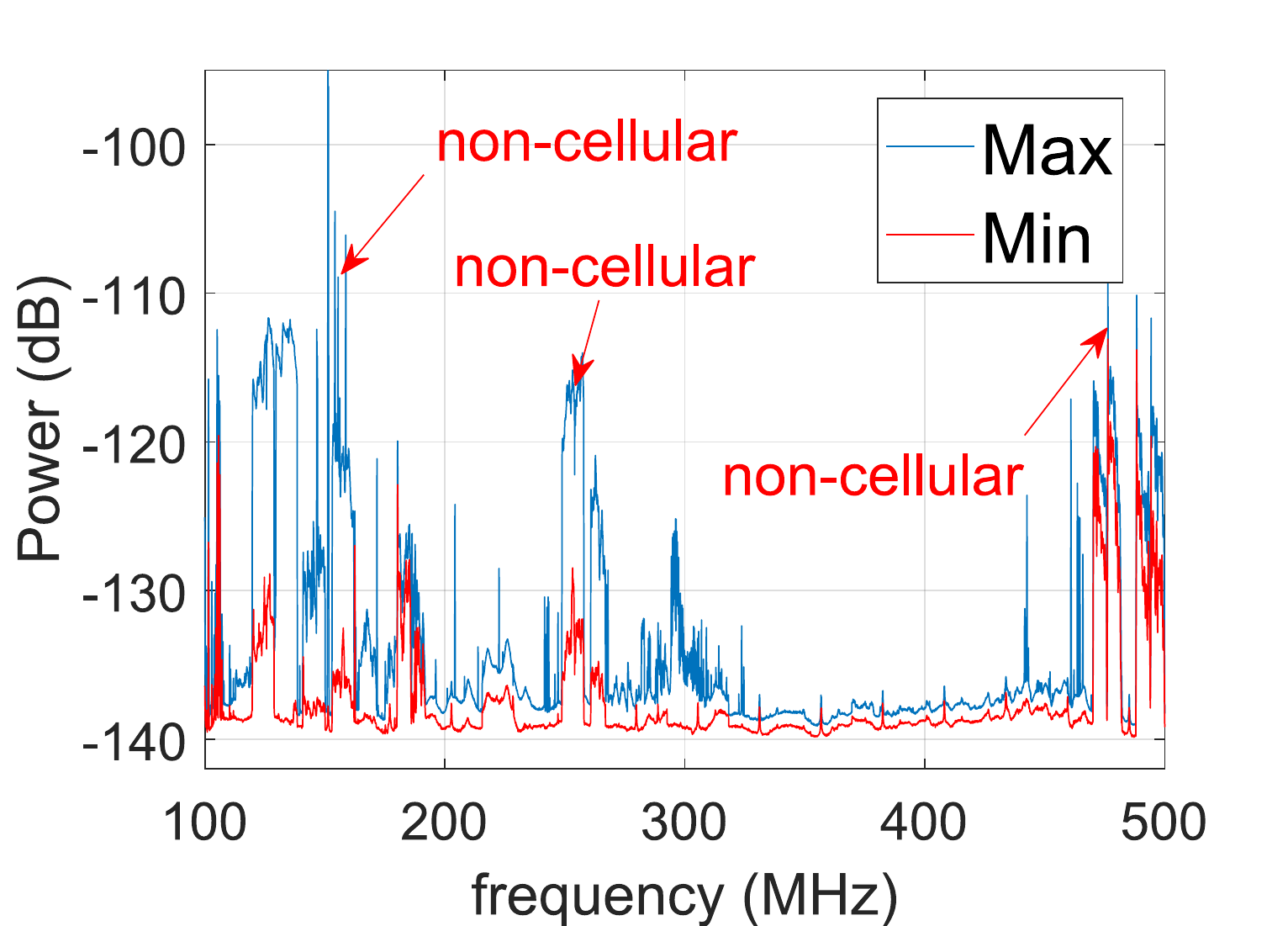}}
    \subfloat[Frequency band 500 - 1000 MHz]{\includegraphics[width=0.5\columnwidth]{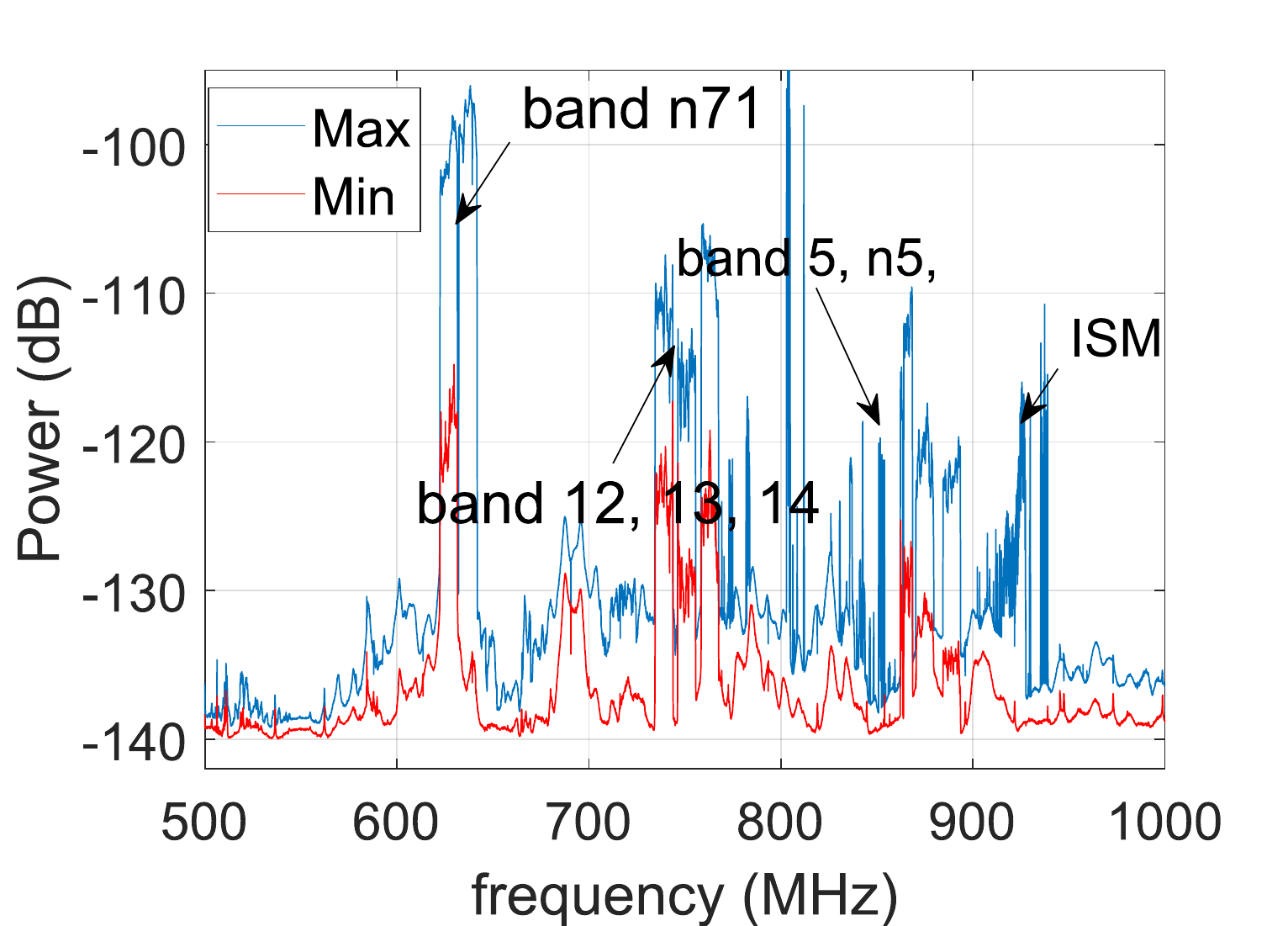}}
    \vspace{-0.1in}
    \subfloat[Frequency band 1 - 1.5 GHz]{\includegraphics[width=0.5\columnwidth]{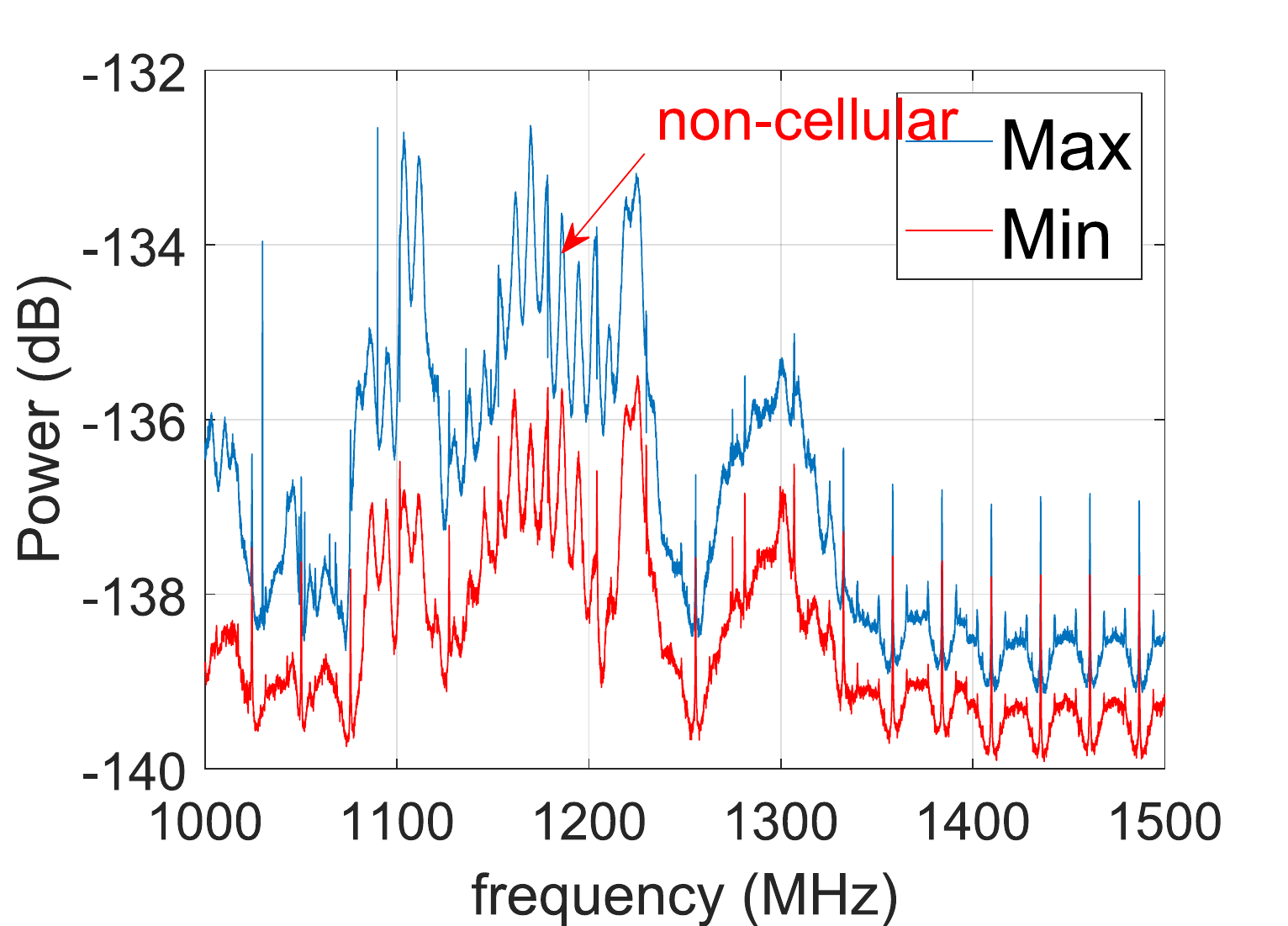}}
    \subfloat[Frequency band 1.5 - 2 GHz]{\includegraphics[width=0.5\columnwidth]{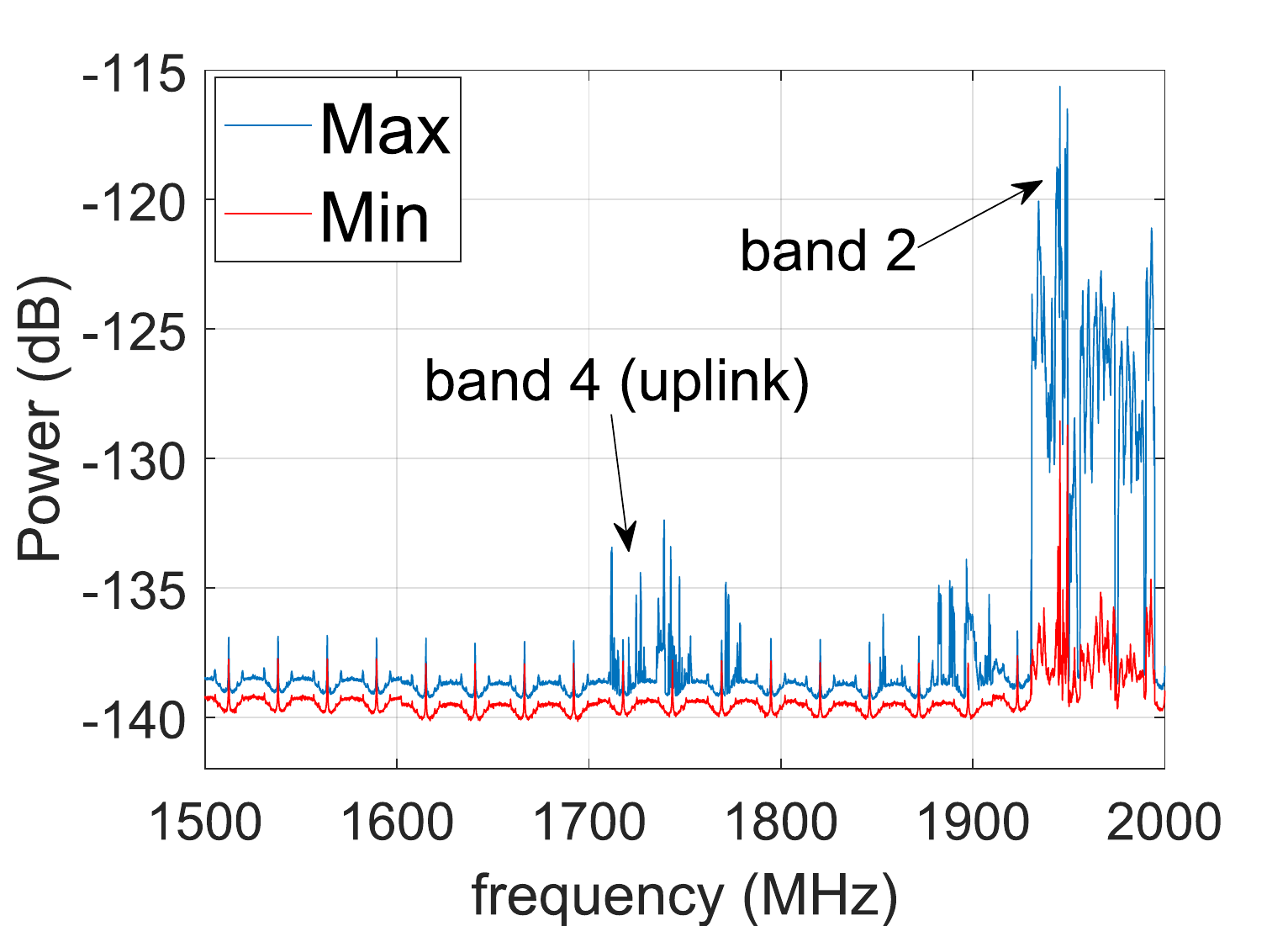}}
    \vspace{-0.1in}
    \subfloat[Frequency band 2 - 2.5 GHz]{\includegraphics[width=0.5\columnwidth]{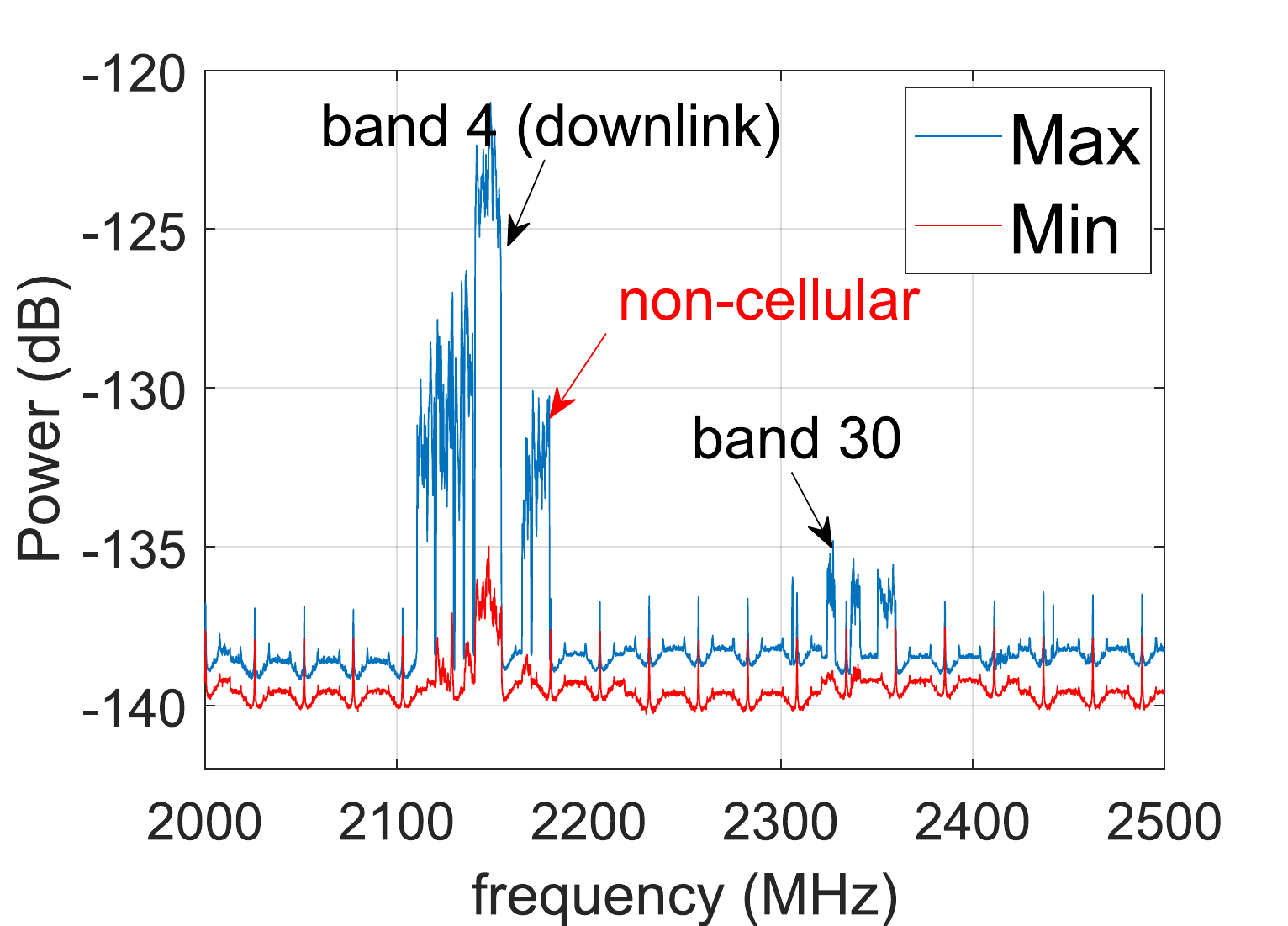}}
    \subfloat[Frequency band 2.5 - 3 GHz]{\includegraphics[width=0.5\columnwidth]{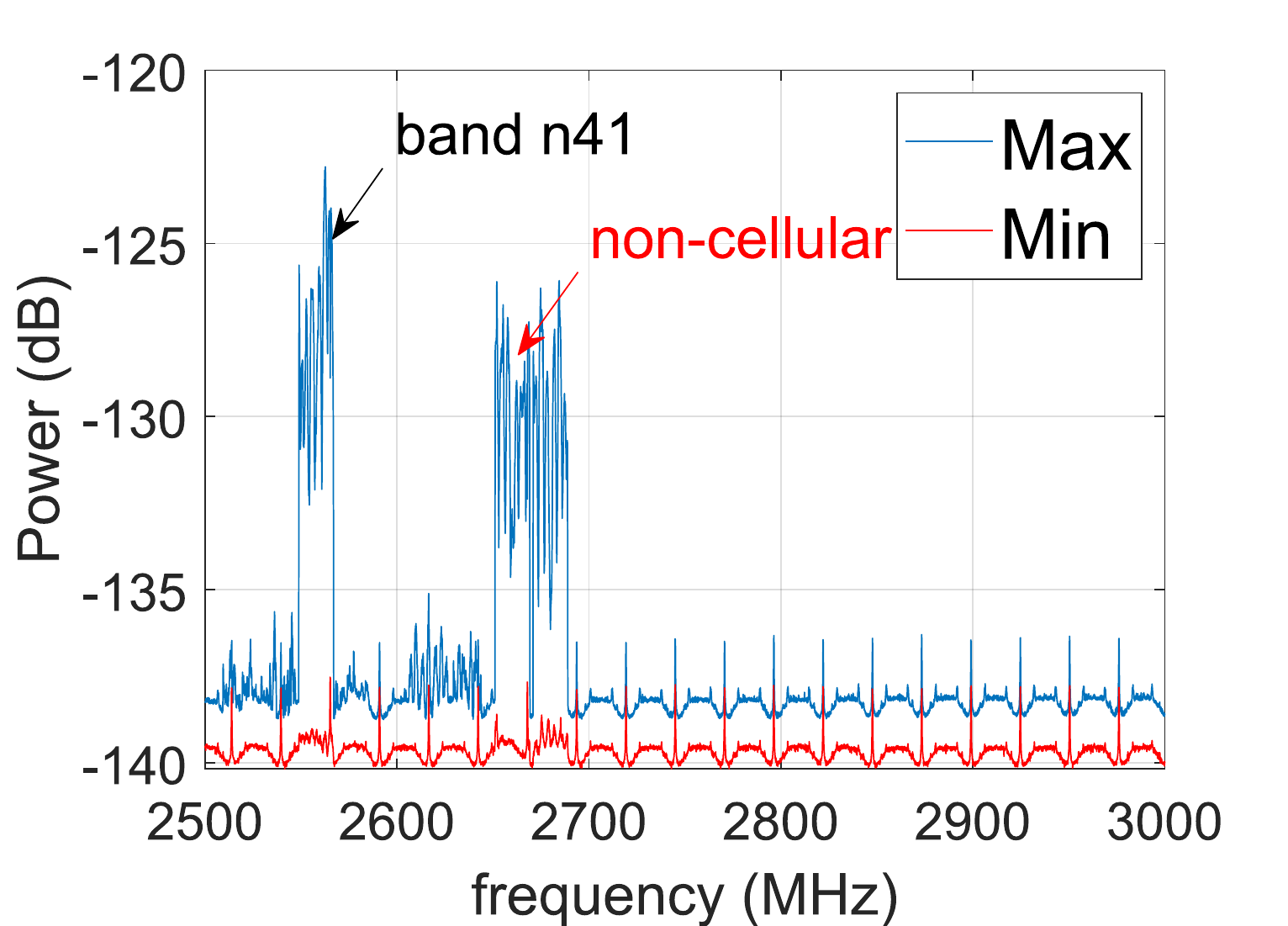}}
	\caption{The enlarged view of the spectrum of the signal power}\label{fig:measurment_subband}\vspace{-0.3in}
\end{figure}

\section{Spectrum Compliance Monitoring}\label{Section:SpectrumCompliance}

In this section, we provide preliminary spectrum measurement results over a period of 24 hours from AERPAW. As discussed earlier, each fixed node in AERPAW includes a spectrum compliance monitoring USRP. The reason for using a spectrum monitoring USRP is that there may be times when an experiment may transmit at a band that it is not authorized to transmit. AERPAW has experimental licenses and an FCC Innovation Zone to use certain bands under some constraints, but transmission at other bands is not allowed. Therefore, AERPAW (or, an NRDZ in a generic sense) needs to detect unauthorized frequency use by an experimenter, and stop such transmissions to be in compliance with FCC regulations. 

Fig.~\ref{fig:spectrum_compliance} shows an illustration of spectrum compliance monitoring, where the sensor is required to distinguish whether the transmissions in a band come from a nearby NRDZ experimenter antenna or from nearby incumbent transmitters. If the transmission in an unauthorized band comes from a nearby AERPAW antenna ($<$ 1 meter apart to the sensing antenna for AERPAW fixed nodes), AERPAW will be required to stop the transmission in that band. The transmission from a close by experimenter AERPAW antenna can be distinguished from transmissions of incumbent users by properly setting a threshold parameter; whenever the received signal strength is larger than a threshold, the transmission is expected to be from the AERPAW experimenter antenna. The threshold can be optimized across different bands and over different times of the day based on past measurements of the spectrum collected at the same location. 
We have collected preliminary data from the spectrum compliance monitoring sensors at the AERPAW fixed nodes. 
Fig.~\ref{fig:measurment_subband} shows spectrum sweeping measurements from 100 MHz to 3 GHz collected at the CC1 fixed node. The mean power and the variance of spectrum observations can be an initial step to calculate the thresholds for identifying out-of-compliance transmissions. 

\begin{table}[t]
\caption{List of United States LTE / NR networks.}
\label{table:LTE_bands}
\centering
\begin{tabular}{p{0.8cm}|p{0.8cm}|p{1.45cm}|p{1.45cm}|p{2cm}}
\hline
Band No & Duplex Mode & Uplink Band (MHz) & Downlink Band (MHz) & Operators \\
\hline\hline
n71 & FDD & 663 - 698 & 617 - 652 & T-Mobile \\
12 & FDD & 698 - 716 & 728 - 746 & AT\&T, T-Mobile \\
13 & FDD & 777 - 787 & 746 - 756 & Verizon \\
14 & FDD & 788 - 798 & 758 - 768 & AT\&T, FirstNet \\
5, n5 & FDD & 824 - 849 & 869 - 894 & AT\&T, T-Mobile, Verizon \\
4 & FDD & 1710 - 1755 & 2110 - 2155 & AT\&T, T-Mobile, Verizon \\
2 & FDD & 1850 - 1910 & 1930 - 1990 & AT\&T, T-Mobile, Verizon \\
30 & FDD & 2305 - 2315 & 2350 - 2360 & AT\&T \\
n41 & TDD & 2496 - 2690 & 2496 - 2690 & T-Mobile \\
\hline\hline
\end{tabular}
\vspace{-0.1in}
\end{table}

\begin{table}[t]
\renewcommand{\arraystretch}{1}
\caption{List of non-cellular frequency allocations \cite{FCC_frequency_table}.}
\label{table:non-LTE_bands}
\centering
\begin{tabular}{p{2.3cm}|p{5.8cm}}
\hline
Name & Frequency Band (MHz) \\
\hline\hline
Aeronautical mobile & 118 - 137, 849 - 851, 894 - 896 \\\hline
Aeronautical radio navigation & 108 - 118, 960 - 1215, 1240 - 1350, 1559 - 1626, 2700 - 2900 \\\hline
Broadcasting (television) & 174 - 216, 470 - 608, 614 - 763, 775 - 793, 805 - 806 \\\hline
Earth exploration satellite & 401 - 403, 1215 - 1300, 2025 - 2110, 2200 - 2290, 2655 - 2700 \\\hline
ISM & 902 - 928, 2400 - 2500 \\\hline
Maritime mobile & 156 - 157, 161 - 163 \\\hline
Maritime radio navigation & 2900 - 3000 \\\hline
Meteorological aids & 400 - 406, 1668 - 1670, 1675 - 1695, 2700 - 2900 \\\hline
Meteorological satellite & 137 - 138, 400 - 403, 462 - 470, 1675 - 1710 \\\hline
Mobile satellite & 137 - 138, 399 - 402, 406, 1525 - 1559, 1610 - 1660,
2000 - 2020, 2180 - 2200, 2483 - 2500 \\ \hline
Radio astronomy & 406 - 410, 608 - 614, 1660 - 1670, 2655 - 2700 \\\hline
Radio determination satellite & 2483 - 2500 \\\hline
Radiolocation & 420 - 450, 902 - 928, 1300 - 1390, 2417 - 2483, 2700 - 3000 \\\hline
Radio navigation-satellite & 1164 - 1240 \\\hline
Space operation & 137 - 138, 400 - 402, 1761 - 1850, 2025 - 2110, 2200 - 2290 \\\hline
Space research & 137 - 138, 400 - 401, 410 - 420, 1215 - 1300, 1400 - 1427,
1660 - 1668, 2025 - 2110, 2200 - 2300, 2655 - 2700 \\
\hline\hline
\end{tabular}
\vspace{-0.2in}
\end{table}

In order to more closely observe the spectrum occupancy at different portions of the spectrum, in  Fig.~\ref{fig:measurment_subband} we zoomed into multiple 500 MHz segments of the spectrum between 100 MHz and 3 GHz. We have also provided a list of common cellular and non-cellular bands used in the United States in Table~\ref{table:LTE_bands} and Table~\ref{table:non-LTE_bands}, respectively. Results in Fig.~\ref{fig:measurment_subband} show particular high cellular activity in the 600-900 MHz segment, as well as the 1.7 and 2.1 GHz bands. As expected, uplink received powers are lower than downlink received powers. In the future we will perform similar measurements across all AERPAW fixed nodes over a longer time period, analyze the data in 3-6 GHz band which had limited activity based on our preliminary measurements. We also aim to collect similar data with AERPAW UAVs and helikites, and compare such data with those collected from ground sensors.     

\section{Conclusion}
In this paper, we present our views on the design, deployment, and operation of NRDZs. We discuss potential research challenges and possible techniques to address them, such as spectrum sensing to build REM, localization and tracking of the interferers, and spectrum compliance monitoring. We discuss experimental research that is needed for  modeling propagation characteristics accurately in an NRDZ, 
, and our plans in integrating new features to the NSF AERPAW platform for  testing and experimentation related to NRDZs. We show related preliminary results on spectrum compliance monitoring, and propose a preliminary analytical framework in power leakage sensing and interpolation for an NRDZ.
\label{sec:conclusion}



\bibliographystyle{IEEEtran}
\bibliography{IEEEabrv,references}

\end{document}